\documentclass{eurocg10}

\begin{document}
\title{ A new separation theorem    with  geometric applications\
}
\author{Farhad Shahrokhi\\
Department of Computer Science and Engineering,  UNT\\
P.O.Box 13886, Denton, TX 76203-3886, USA\
farhad@cs.unt.edu
}

\date{}
\maketitle
\date{} \maketitle

\begin{abstract}

Let $G=(V(G), E(G))$ be an undirected  graph with  a  measure function
$\mu$ assigning non-negative values to  subgraphs $H$ so that  $\mu(H)$
does not exceed the clique cover number of $H$. 
When $\mu$ satisfies  some additional  natural conditions, 
we study the problem of separating $G$ into two subgraphs,  
each with a measure  of at most  $2\mu(G)/3$ 
by removing a set of vertices that can be covered with a small
number of cliques $G$.
When $E(G)=E(G_1)\cap E(G_2)$, 
where $G_1=(V(G_1),E(G_1))$ is  a  graph with $V(G_1)=V(G)$, 
and  $G_2=(V(G_2), E(G_2))$  is a chordal graph with $V(G_2)=V(G)$,
we prove that there is a  separator $S$  that can be covered   
with $O(\sqrt{l\mu(G)})$ cliques in $G$, where $l=l(G,G_1)$ is a  parameter 
similar to the bandwidth, which   arises   from the  linear orderings
of cliques covers in  $G_1$.  
The  results and the methods  
are  then used to  obtain   exact and approximate algorithms  
which significantly improve  some of the past results  for 
several  well  known NP-hard geometric  problems.
In addition, the methods involve introducing  new concepts  
and hence may be of an  independent interest.

\end{abstract}

\section{Introduction and Summary}
Separation  theorems have shown to play a key role in the design of 
the divide and conquer  algorithms, as well as 
solving  extremal  problems  in combinatorial  topology and geometry.
The earliest result in this area  is a  result of Lipton and Tarjan
\cite{LT} that asserts  any $n$ vertex   planar  graph  can be
separated into  two subgraphs with at most $2n\over 3$ vertices
by removing only  $O(\sqrt{n})$ vertices. This result is extended
by many authors including  Miller et al \cite{Mi},
Fox and Pach \cite{FP1}, \cite{FP2},  
and Chan \cite{Ch}.

Clearly if a graph contain a large  clique, then  it can not have 
a separation property  that resembles the planar case.
Fox  and Pach \cite{FP1,FP2} have recently studied
the string graphs which contain  the class of planar graphs,    
and have shown that when  these  graphs do not contain
 a $K_{t,t}$, of fixed 
size $t$, as a subgraph,  then a suitable separator exits. Although this  
powerful result is extremely effective  in solving extremal problems,
its computational power is limited to graphs that 
do not contain a  "large" complete  bipartite subgraph. 
Chan \cite{Ch} studied the problem of computing the packing and piercing
numbers of fat objects in $R^d$, where the dimension $d$ is fixed.   
He  drastically  improved    the  
running time  of the   first  polynomial time approximation scheme  (PTAS) 
for packing of fat objects due to Erlebach et al \cite{Erl}, and also
provided the first PTAS for the piercing problem of fat objects.   
Parts of Chan's \cite{Ch}  work involved
proving a separation  theorem   with respect to the  abstract concept of a  
measure  on fat objects.  Motivated by his work  we have  defined  
the notation of a measure  in a more combinatorial fashion on graphs.
Furthermore, we  have  proven a combinatorial separation theorem.
It should however  be noted that   the  results in \cite{Ch} do not imply
ours, and our  results do not apply to the  general fat objects.

Let $\mu$ be  a function that assigns non-negative values
to subgraphs of $G$. 
$\mu$ is called a  measure function if the following hold.

$(i)~$ $\mu(H_1)\le \mu(H_2)$, if $H_1\subseteq H_2\subseteq G$,
$(ii)~$ $\mu(H_1\cup H_2)\le \mu(H_1)+\mu(H_2)$, 
if  $H_1,H_2\subseteq G$, 
$(iii)$  $\mu(H_1\cup H_2)=\mu(H_1)+\mu(H_2)$, if there are no edges between $H_1$ and $H_2$.

Central to our result is a length concept similar to the  bandwidth.
Let $H$ be a graph  with $V(G)=V(H)$ and $E(G)\subseteq E(H)$
and let $C=\{C_1,C_2,...,C_k\}$ be a clique cover in  $H$. For any $e=xy\in E(G), x\in C_l,y\in C_t$, define  $l(e,G,H,C)$ to be $|t-l|$.
Let  $l(G,H,C)$ denote   $\max_{e\in E(G)}l(e,G,H,C)$, and 
let $l(G,H)$ denote $\min_{C\in {\cal C}} l(G,H,C)$, where ${\cal C}$ denotes
the set of all ordered clique covers in $H$. We refer to  $l(G,H)$ as the 
{\it length} of $G$ in $H$.    
It is important  to note that  when
$l(G,H)$ is small, then $G$ exhibits some nice separation properties. 
For instance, one can partition  $V(G)$ into blocks of $l(G,H)$
consecutive cliques of $H$, and argue that removal of any block separates
$G$.
Particularly, when $l(G,H)=1$, then one can separate $G$ by removing one  
clique form $H$. Similar important  concepts such as  treewidth,
pathwidth, and  bandwidth have been introduced in the past \cite{Bo},
but none is identical to the  concept  of  length introduced here.
Clearly $l(G,G)\le BW(G)$, 
where $BW(G)$ is the bandwidth of $G$. Moreover as  we will see, there is
a simple but important connection between  $L(G,G)$ and the
dimension of interval orders.\\

Recall that a chordal graph does not
have a chordless cycle of length at least 4. 
Our main result which is Theorem \ref{t1}  is a generalization of the result 
stated earlier in the abstract, to $p\ge 2$  graphs,  
where $G_p$ is a chordal graph.  

\begin{theorem}\label{t1}

{\sl Let $\mu$ be a measure on  $G=(V(G),E(G))$, and  let $G_1,G_2,...,G_p$ 
 be  graphs  with $V(G_1)=V(G_2)=,...,V(G_p)=V(G)$, $p\ge 2$ and 
$E(G)=\cap_{i=1}^p E(G_i)$ so that $G_p$ is chordal. Then there is 
a vertex separator $S$ in $G$ whose removal separates $G$
into two subgraph so that each subgraph has a measure of at most 
 $2\mu(G)/3$.  In addition, 
the induced graph of $G$ on $S$   can be covered with 
at most $O(2^p{l^*}^{p-1\over p}{\mu(G)}^{p-1\over p})$ many  cliques 
from $G$, where $l^*=max_{1\le i\le p-1}l(G,G_i)$.
}

\end{theorem}

Proof of Theorem \ref{t1} combines   the clique separation 
properties of chordal graphs    and   perfect elimination trees,
together with  the properties associated with the length  of a  graph.
The theorem either finds a suitable  clique separator in the 
chordal graph $G_p$, or identifies a graph $G_j$, for which 
the cardinality of the clique cover is large,  and
the  separates $G$ using length properties. 
An effective  application of Theorem \ref{t1} to a  specific problem 
normally requires to define  $\mu$ so that,  $\mu(G)=O(|C|)$, for an appropriate
 clique cover $C$ in $G$.

The time complexity of finding the separator depends on the structure
of the  measure and  how fast
we can compute the  measure on any subgraph. In typical
applications of interest with $p=2$, the separation  algorithm 
can be implemented to run  better than
 $O(|V(G)|^2)$. \\ 

\section{Applications}
Proper applications of Theorem \ref{t1} gives rise to the following.

\begin{theorem}\label{t2}

{\sl Let  $G=(V(G),E(G))$ be the intersection graph of a  
set of axis parallel unit height rectangles in the plane. 
Then, a maximum independent set in $G$ can be computed in 
${|(V(G)|}^{O({\sqrt{\alpha(G)}})}$, where $\alpha(G)$ is 
the independence number of $G$. Moreover, there is a
PTAS that gives a $(1-\epsilon)-$approximate solution 
to $\alpha(G)$ in 
${|V(G)|}^{O({1\over\epsilon})}$ time
and requires  $O({|V(G)|})^2$ storage.
}

\end{theorem}

{\bf Proof sketch.}
For $R_1,R_2\in V(G)$, define $R_1\prec_1 R_2$, if there
is a horizontal line $L$ so that $R_1$ is above $L$ and $R_2$ is below $L$.
Likewise, define $R_1\prec_2 R_2$, if there is a  vertical line   
$L$ so that $R_1$ is to the left of $L$ and $R_2$ is to the right
of $L$. Observe that  $G_i$,  the incomparability graphs  for $\prec_i$
is an interval  graph, $i=1,2$,  and hence is chordal. It is further
easy to verify that $l(G,G_1)=1$.  Finally, let $C$ be
a $2-$approximate solution for the clique cover number of $G$ that 
also provides for a $1/2-$approximate solution to independence number
of $G$, and for any induced subgraph $F$, define $\mu(F)$ to be the  restriction of $C$ to $F$.
(Note that $\mu(G)$ can be computed in $O(|V(G)|\log (|V(G)|))$
time \cite{Ch2}.)\\

For obtaining the sub-exponential time algorithm one can adopt  the 
method in  \cite{LT} proposed for planar graphs,  by  enumerating
independent sets inside of separator and then   recursively applying
Theorem \ref{t1} to $G$.  For the PTAS, one can also use the 
original approach in \cite{LT} adopted by Chan \cite{Ch}, by  recursively
separating $G$, but terminating  the recursion when  
for a subgraph $F$, $\mu(F)=O({1\over \epsilon}^2)$ and then   applying the
sub-exponential algorithm to $F$.
$\Box$\\

Similarly,  one can prove the following.

\begin{theorem}\label{p2}

{\sl Let  $S$ be  a set of  axis parallel
unit height rectangles in the plane. 
Then, the piercing number of $S$  can be computed in 
${|S|}^{O({\sqrt{P(S)}})}$, where $P(S)$ is 
the piecing number of  of $S$. Moreover, there is a
PTAS that gives a $(1+\epsilon)-$approximate solution 
to $P(S)$ in 
${|S|}^{O({1\over\epsilon})}$ time
and requires  $O({|S|})^2$ storage.
}

\end{theorem}

Our sub-exponential time algorithms in Theorems \ref{t2} and \ref{p2} are 
the first
ones for the unit hight rectangles, and we are not aware  of any previous 
sub-exponential algorithms for these problem. 
Moreover, the storage requirement for the PTAS in  Theorem \ref{t2} drastically  improves upon
${|V(G)|}^{O({1\over\epsilon})}$  storage requirement
of  the best known previous algorithm in \cite{AG},  due to Agarwal et al, 
that was combining dynamic programming  with the shifting method.  
Finally the time complexity of PTAS in Theorem \ref{p2}  
drastically  improves upon
${|S|}^{O({1\over\epsilon}^2)}$ in \cite{Ch2}.\\

\begin{theorem}\label{t3}

{\sl Let $P, |P|=n$ be a set  of points in the plane.
There is an algorithm for computing  the minimum number of discs of unit
diameter    needed to cover all points in $P$  that gives an 
answer in $n^{O(\sqrt{opt(P)}~)}$ time, and $O(n^2)$ storage
where $opt(P)$ is the  value of  an optimal solution.
Furthermore,   there is a PTAS that gives a $(1+\epsilon)-$approximate solution 
in 
$n^{O({1\over \epsilon})}$ time, and $O(n^2)$
storage.}

\end{theorem}

{\bf Proof  sketch.}
For graph $G$, let $V(G)=P$, and for any $x,y\in P$, if they
of distance at most 1, then place $xy\in E(G)$. 
Next,  as suggested in \cite{Hoch} consider  a square $n$ by $n$  grid 
in the plane   containing all the points,   so that each cell  in the grid is 
a unit square. Note that the grid can be placed  so that no two
points appear in the boundary of a cell.
Define two   interval orders  $\prec_1$ and $\prec_2$ on $V(G)$ as follows.   
$x\prec_1 y$, if $xy\notin E(G)$  and $x$ and $y$ are in different
vertical strips of the grid so that $x$ is to the left of $y$. 
$x\prec_2 y$, if $xy\notin E(G)$ and there is horizontal line $L$ in the plane
so that $x$ is above $L$ and $y$ is below $L$. 
Let $G_i,$ $i=1,2$ be the incomparability interval
graph associated with $\prec_i$,  and note that points in any vertical strip 
of the grid constitute a clique of $G_1$ and hence $l(G,G_1)=1.$\\

For any $xy\in E(G)$, $x,y\in E(G)$, 
place two  discs in the plane  
that has $x$ and $y$ in its boundary  and call the resulting multi-set of discs
$\cal C$, and note that $|{\cal C}|=O(n^2)$.  
If $G$ is disconnected, then  we would  solve the problem for each component,
and take the union of the solutions, so we will assume that $G$ is connected.
Thus, we can  assume with no loss of generality that 
any  feasible solution $C'$ for any $P'\subseteq P$  is a subset of $\cal C$, for otherwise we 
can replace    any  $D\in C'$ by one  disc from $\cal C$.  
Furthermore, it is easy to  construct  a feasible solution $C$ so that  
$|C|\le c\beta(G)$, where $\beta(G)$ is the clique cover number of $G$ and $c$ is a  constant no more than 16,  in about  $O(n^2)$ time.
Thus $\beta(G)\le opt(P)\le |C|\le 16\beta(G)$.
For any induced subgraph $F$ in $G$, define $\mu(F)$ to be 
$|C_{F}|$, or, the cardinality of the   restriction of $C$  to $F$,
and note that   Theorem \ref{t1}
applies.

Finally follow the details in \cite{LT} and  the 
previous theorems by noting that we will always select our disc cover  
solutions from $\cal C$.
(Note that the time complexity of enumeration inside of the   
separator $n^{O(\sqrt{opt(P)})}$.) $\Box$\\

Our    sub-exponential time 
algorithm in Theorem \ref{t3} is the first one  for the covering problem
of Hochbaum and Mass \cite{Hoch}. In addition  the time complexity of 
our PTAS  drastically  improves 
 time complexity of the original algorithm that was   
${n}^{O({1\over\epsilon}^2)}$ in \cite{Hoch}.\\

Let $\prec$ be a partial order on a finite set $S$. The  dimension of $\prec$,
denoted by $dim(\prec)$, is the minimum number of total orders on $S$ 
whose intersection is $\prec$ \cite{To}.
We finish this section by stating  a  simple theorem  that establishes
some connections between partial orders, the dimension of interval orders and 
the concept of length introduced here.\\

\begin{theorem}\label{t4}

{\sl Let $G$ be a  graph.

$(i)~$ If $l(G,G)=1$, then $G$ is an incomparability graph.

$(ii)~$ If $G$ is an interval graph  whose underlying interval order
is $\prec$, then $dim(\prec)\le l(G,G)+2$.

}

\end{theorem}

{\bf Justification.} For $(i)$, let $(C_1,C_2,...,C_k)$ be a clique cover of
$G$ so that for $x\in C_i$ and $y\in C_j$, we have $xy\in E(G)$ only
if $|i-j|\le 1$. Now for any $x,y\in V(G)$ with $x\in C_i$, $y\in C_j$
with $j>i$, define $x\prec y$, if and only, if  $xy\notin E(G)$.
It is easily seen that $\prec$ is partial order on $V(G)$ so that
$\bar G$ or the  complement of $G$ is a comparability graph, and hence 
$G$ is an incomparability graph.  We omit the  proof of $(ii)$.  
$\Box$\\

{\bf Remarks: } A proof of 
Theorem \ref{t4} (ii) has appeared in Congressus Numerantium 205 (2010), 105-111.
Additionally, a small number of minor corrections were made to this paper
after its publications in EuroCG2010.

\end{document}